\documentclass[final,3p,times,twocolumn]{elsarticle}
\usepackage{amssymb}
\usepackage{soul}
\usepackage{amsmath}
\usepackage{float}
\usepackage{xcolor}
\definecolor{RED}{RGB}{156,78,90}

\journal{International Journal of Heat and Fluid Flow}

\usepackage{color}
\definecolor{B}{RGB}{52,78,200}

\begin{document}
\begin{sloppypar}
\begin{frontmatter}

\title{Bubble entrainment in turbulent jets leaping from liquid surface}

\author[1]{Fangye Lin}
\address[1]{Ningbo Innovation Center, Zhejiang University, Ningbo 315100, China; State Key Laboratory of Fluid Power and Mechatronic Systems, Zhejiang University, Hangzhou 310027, China}

\author[2]{Mingbo Li\corref{cor1}}
\ead{mingboli@sjtu.edu.cn}
\cortext[cor1]{Corresponding author}
\address[2]{Key Laboratory of Hydrodynamics (Ministry of Education), School of Ocean and Civil Engineering, Shanghai Jiao Tong University, Shanghai 200240, China}

\begin{abstract}
We investigate the phenomenon of air entrainment in turbulent water jets exiting a pool near the free surface. Our experimental results reveal that bubble entrainment occurs only within a specific region close to the point where the jet exits the water and is dictated solely by the jet's exit velocity, rather than the Reynolds number. The morphology of the jet above the pool surface, influenced predominantly by the Froude number or the injection angle, classifies the flow into two regimes: curtain jet and column jet. However, variations in jet morphology have minimal impact on the critical velocity required for bubble entrainment. Our findings suggest that bubble entrainment is driven by the dynamic interplay of shear forces and instabilities. As the jet exits the nozzle, it interacts with the surrounding fluids, amplifying instabilities through Kelvin-Helmholtz mechanisms. These disturbances generate intense fluctuations on the jet surface, creating localized low-pressure zones that trap air and entrain bubbles. As the jet progresses further into the air, capillary forces dampen surface instabilities, diminishing the jet’s capacity to sustain bubble entrainment at longer distances. This study offers new insights into the mechanics of air entrainment in turbulent water jets, emphasizing the role of injection velocity and instabilities in the entrainment process. 
\end{abstract}


\end{frontmatter}

\section{Introduction}

Air entrainment is a crucial factor in a wide range of agricultural and industrial processes, particularly aiding in the transportation of oxygen in aquaculture and wastewater treatment \cite{alkahalidi2015waterej,temesgen2017advcolint}. However, it's important to note that entrapped air bubbles can also have adverse effects, such as generating acoustic noise and causing cavitation damage in hydraulic systems \cite{plesset1977annulrev}. The process of air entrainment typically begins with the deformation of the free surface, leading to the creation of concave surfaces that subsequently trap the air bubbles\cite{zhang2019bubble}.

In a variety of scenarios, entrainment has been identified as a significant phenomenon. For instance, when a mass falls and impacts the free surface, entrainment occurs \cite{oguz1995jfm}. In the case of a droplet striking a liquid pool, the convergence of capillary waves on the free surface can result in the trapping of millimeter-sized air bubbles \cite{oguz1990jfm,pumphrey1990jfm}. Moreover, the formation of concentrated vortex rings in some cases leads to the entrainment of large bubbles, typically centimeter-sized \cite{thoraval2016pre,deka2017pof}. When a solid body impacts the free surface, a cavity grows deep and eventually seals in the middle of the wall due to surface deformation induced by gravity \cite{bergmann2016prl}. 

Air entrainment is also evident for a range of jet conditions~\cite{li2019jet, eggers2008physics}, especially when a jet plunges into a pool~\cite{kiger2012annulrev,eggers2008rpp}. A liquid column moving through a gas before impacting a liquid pool is known as a plunging jet, which has been widely studied since the 1970s~\cite{lin1966gas, van1973surface}. Air entrainment in this scenario depends on several factors, including the fluid's physical properties, the geometric characteristics of the jet system (such as nozzle diameter, height of the jet fall, nozzle shape, and angle), and operational conditions like fluid velocity, temperature, and pressure~\cite{bin1993gas, detsch1990critical, baylar2006development, chirichella2002incipient, gomez2011impact}. The numerous experimental results confirmed that the differences in the entrainment process depend on the jet velocity. The air entrainment process varies for jets with low impact velocity (less than $\sim$4 m/s) compared to those with higher velocities (4 $\sim$ 12 m/s). Despite this variation, the general sequence of events during entrainment remains similar for both velocity ranges. When the jet velocity reaches a critical value, known as the onset velocity~\cite{miwa2019chemet,lins1966aiche}, the pool's surface can no longer keep pace with the oscillations. The pool surface concaves downward, creating an air layer underneath the pool level around the plunging jet~\cite{eggers2001prl}. This pocket creates a void through which air is drawn into the pool. The interface of the air pocket is inherently unstable, primarily due to the presence of a re-entrant jet that seeks to fill the void~\cite{zhu2000jfm}. This instability causes the air pocket to break up into numerous smaller bubbles. In high-velocity jets, this mechanism also occurs, but the air pocket fluctuates more significantly compared to lower velocity jets. Once entrained, the bubbles are transported within the pool by large-scale eddies, which are typically oriented perpendicular to the flow direction. This observation highlights the influence of turbulence on the behavior of entrained bubbles and their movement within the pool. Apart from interactions between a free surface and impacting mass, air entrainment is commonly observed when turbulent water jets discharge into the atmosphere~\cite{chanson1996book}. Irregular vortex action and turbulent velocity fluctuations on the free surface are believed to contribute to air entrainment.

In this study, air entrainment is also observed when submerged water jets exit a pool. This phenomenon is a fundamental fluid dynamic process found both in nature and in everyday life, such as archerfish using water jets to shoot down insects above the water surface \cite{vailati2012plosone}, or in the use of underwater fountains for firefighting and landscape design. Previous research \cite{zou2017pof} on viscous (laminar) jets has demonstrated that as a jet moves through a submerged layer of liquid, it can transport and eject a significant amount of liquid above the water surface. Additionally, the presence of the free liquid surface plays a significant role in shaping the jet.

At lower velocities, the jet is primarily influenced by capillary forces, forming a water curtain-like structure above the free surface. However, when the velocity surpasses a critical threshold, inertia becomes the dominant force, transforming the jet into a cylindrical shape. The amount of liquid entrained from the pool and the resulting jet morphology are influenced by several factors, including the injection velocity, inclination angle, needle inner diameter, immersion depth, and immersion distance. Despite these dynamics, no bubbles were found to be entrained when the jet broke through the free surface. In this work, we increased the velocity of the water jet to achieve turbulent conditions as it exited the pool, conducting a detailed investigation into how various injection parameters affect bubble entrainment. This study aims to provide insight into the mechanisms by which bubbles are trapped during the jet’s exit from the water surface.

\section{Experiments}

\begin{figure}
\centering
\includegraphics[width=0.47\textwidth]{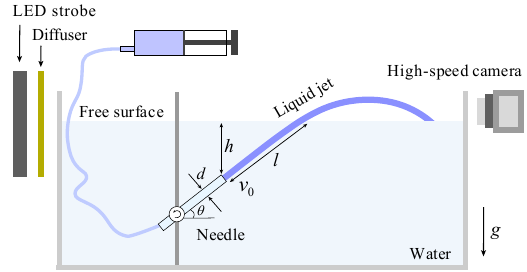}
\caption{Sketch of the experimental set-up with geometrical parameters defined.}
\label{FIG1}
\end{figure}

The experimental setup consisted of a rectangular Plexiglas tank measuring $600 \times 150 \times 150\ \rm{mm}^3$, filled with deionized water (with a depth of 100 mm), and a syringe pump that generated a water jet through a stainless-steel needle, as depicted in Figure \ref{FIG1}. In this study, five types of needles with inner diameters $d$ of 0.90, 1.25, 1.45, 1.69, and 2.40 mm were utilized. The mean exit velocity of the jet, denoted as $v_0$ (1.21$\sim$5.57 m/s), is varied to change the the jet Reynolds number, 
\begin{equation}
\label{eqn:P}
{\rm{Re}} = \frac{\rho v_0 d}{\mu}
\end{equation}
in the range $1900\sim8600$ with $\rm \rho = 998\ kg/m^3$ the liquid density, $\rm \mu = 0.89\ mPa \cdot s$ the dynamic viscosity of water. 

\begin{table}
\caption{\label{tab:table1}Principal dimensions of the experimental setups.}
\begin{tabular}{cccccccc}
\hline\hline
Parameter&Definition &Values\\
\hline 
Injection velocity& $v_0$ (m/s) & 1.21 $\sim$ 5.57\\
Inclined angle of needle & $\theta$ ($^{\circ}$) & 30 $\sim$ 70\\
Inner diameter of needle & $d$ (mm) & 0.90 $\sim$ 2.40\\
Immersion depth & $h$ (mm) & 1.00 $\sim$ 5.39\\
Immersion distance & $l$ (mm) & 1.57 $\sim$ 7.26\\
\hline\hline
\end{tabular}
\end{table}

The needle was submerged in the water and held at an inclined angle $\theta$ using a clamp, which was affixed to a manual displacement platform. The platform allowed for adjustment of the needle's immersed depth, denoted as $h$ (the distance from the needle's tip to the free surface), within the range of 1.00 to 5.39 mm, with a positional accuracy of 0.01 mm. To determine the position of the free surface ($h = 0$), the needle was initially submerged and then gradually raised by the manual displacement platform until its tip approached its mirror image~\cite{zou2017pof}. The distance that the jet traveled from the nozzle to the free surface was given by $l = (h/\sin \theta) + (d/2\tan\theta)$. These physical parameters and ranges of the experiments are summarized in Table ~\ref{tab:table1}. To visualize the flow field evolution of the jet, additional experiments were conducted, marking the liquid in the syringe with red ink at a concentration of less than 0.5$\%$. The behavior of the jet was recorded from the side view using a monochrome high-speed camera (Phantom V2512) at a frame rate of 4000 fps and a colored camera (Photron FASTCAM Mini AX) at a frame rate of 2000 fps. All experiments were conducted at room temperature (25$^{\circ}$C).

\section{Results and discussion}

\subsection{Regimes of the jet morphology}

\begin{figure}[!t]
\centering
\includegraphics[width = 0.46\textwidth]{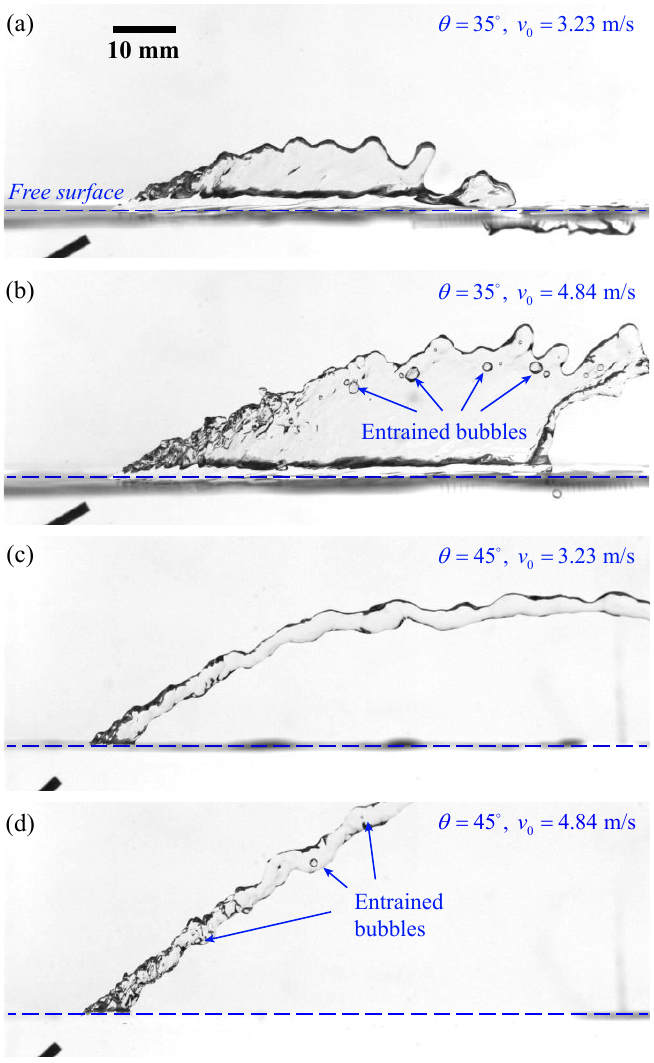} 
\caption{Four kinds of jet regimes above the free surface: (a) Water curtain without bubble entrainment (Multimedia available online), (b) Water curtain with bubble entrainment (Multimedia available online), (c) Water column without bubble entrainment (Multimedia available online), and (d) Water column with bubble entrainment (Multimedia available online). The inner diameter of the needle is $d = 1.4$ mm. The immersion depth is $h = 6.00$ mm.}
\label{fig2}
\end{figure}

Typical flow structures of water jets emerging from a pool are illustrated in Figure \ref{fig2}. These flow patterns resemble those observed in viscous jets ($\rm{Re} < 1000$) exiting a liquid pool, as described by Zou et al.~\cite{zou2017pof} Based on the characteristics of the jet above the free surface, two distinct regimes can be identified: the curtain-jet regime (Figure \ref{fig2}(a) and (b), Multimedia available online) and the column-jet regime (Figure \ref{fig2}(c) and (d), Multimedia available online).

When the jet exits the pool at a low incline angle ($\theta = 35^{\circ}$), the upper edge of the jet curves while the lower edge remains close to the water's surface, forming a liquid curtain that hovers just above the water level. The water jet maintains close proximity to the free surface, and the jet's upper edge bends, creating a "curtain" effect. The interaction causes the formation of a thin liquid sheet (curtain) just above the water level. The curved upper edge is due to the balance between inertial forces driving the jet and gravitational forces pulling the water back down.

In this regime, due to the proximity of the jet to the water surface, air resistance and surface tension stabilize the lower part of the jet, keeping it attached to the free surface. The relatively lower velocity in this regime means that the disturbances, while present, are not as prominent as in the column regime. However, as the jet exits, small perturbations may grow, particularly if air entrainment occurs, leading to localized disturbances that travel along the jet. These disturbances often appear as ripples or waves that increase in size as they propagate downstream. Eventually, this leads to the breakup of the liquid curtain into smaller droplets and finger-like structures at its edges. 

In contrast, at a steeper incline angle ($\theta = 45^{\circ}$), the lower edge of the jet detaches from the water's surface, creating a column-like flow structure. The jet’s lower edge separates from the free surface, allowing it to form a distinct, column-like structure. In this regime, the velocity of the jet is typically higher, which induces more intense turbulent interactions. As the jet becomes fully detached, it no longer interacts directly with the pool’s surface, and the column is suspended in the air, dominated by inertial forces rather than surface tension. 

In comparison to the "curtain jet of viscous liquid" and "column jet of viscous liquid", which are characterized by steady structures and smooth rims~\cite{zou2017pof}, the turbulent water curtain- and column-jets exhibit significant disturbances near the exit point. When the jet velocity is high enough, air bubbles can be entrained into the curtain near the point of exit (Figure~\ref{fig2}(b) and (d)). These bubbles then travel along with the jet, leading to the formation of a self-aerating jet. As the jet moves further, the disturbances increase in both wavelength and amplitude, leading to a finger-like rim and the eventual breakup of the jet. The progression of these disturbances inhibits further bubble entrainment, meaning that few new bubbles are observed in regions farther from the exit point. 

\begin{figure}[!t]
\centering
\includegraphics[width = 0.47\textwidth]{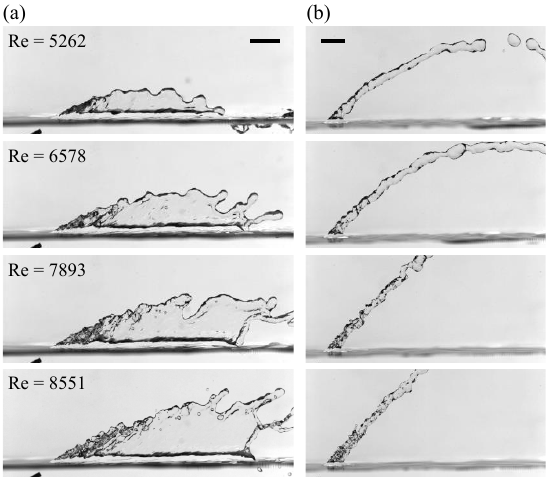}
\caption{Effect of jet Re on the steady jet structures for (a) curtain-jet regime ($\theta = 30^{\circ}$, $h = 2.37$ mm) and (b) column-jet regime ($\theta = 50^{\circ}$, $h = 4.13$ mm). Scale bar for both cases represents 10 mm. }
\label{vortex111}
\end{figure}

In Figure~\ref{vortex111}, the curtain- and column- jet structures produced with different Reynolds numbers are shown. With the increase of jet Reynolds number, more water is taken out of the pool and a larger water curtain is formed (Figure~\ref{vortex111}(a)). In particular, the expansion of the turbulent mixing area causes more bubbles to be brought in. At high Reynolds numbers, the tail of the curtain jet has become extremely unstable and is easily torn under disturbance. The amplified instabilities lead to the eventual breakup of the jet, a process often modeled by Rayleigh-Plateau (R-P) instability \cite{Rayleigh1879breakup,Eggers1997breakup}, where the cylindrical column of liquid disintegrates into droplets due to the destabilizing effects of surface tension and inertia. Bubble entrainment mainly occurs at the upper edge of the jet outlet. For column-jet cases with a very weak disturbance level (usually no readily visible perturbation to the jet surface), the surface of the jet is smooth (Figure~\ref{vortex111}(b)). If given sufficient time, it will break up into a series of discrete drops or lumps under the interference of long wave. At higher Re, the surface of the jet becomes rough with increasing disturbance level, especially at the exit. For highly disturbed impacting jets, the jet becomes nonuniform, characterized by a visibly rough surface with wavelengths varying from those dictated by capillary-shear balance up to the size of the jet. 

\begin{figure}[!t]
\centering
\includegraphics[width = 0.468\textwidth]{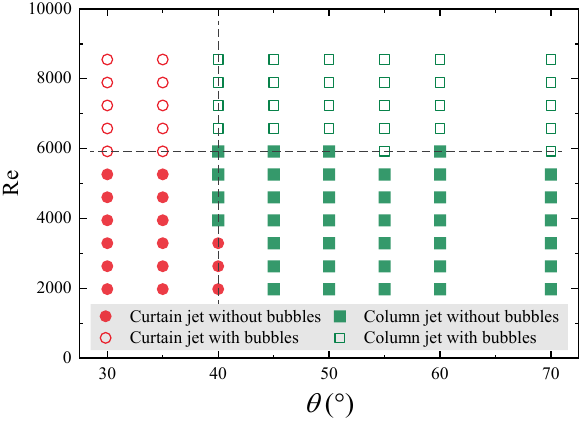}
\caption{Phase diagram for the jet pattern with respect to the angle of inclination $\theta$ and Re.}
\label{regime}
\end{figure}

\begin{figure*}[!t]
\centering
\includegraphics[width = 0.97\textwidth]{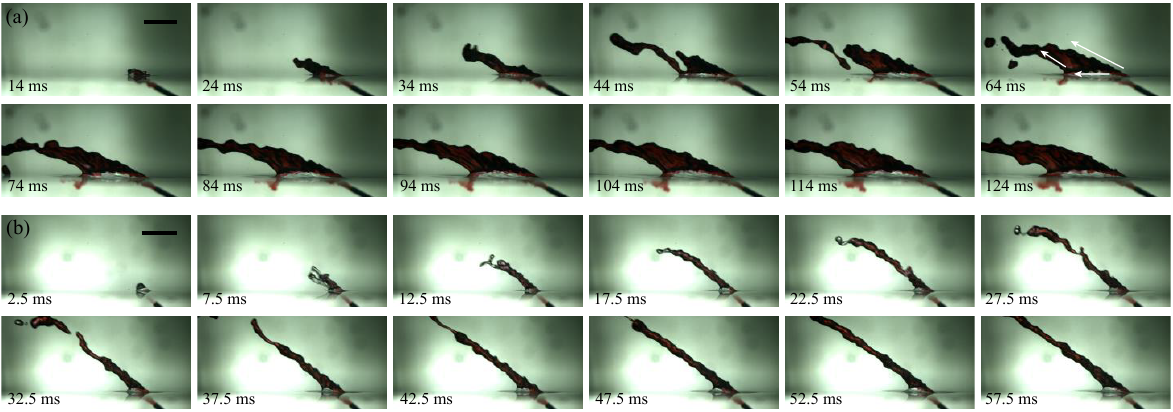}
\caption{Temporal evolution of the leaping jet for (a) curtain-jet regime (The diameter of the needle $d = 0.9$ mm, the inclined angle of needle $\theta = 30^{\circ}$, the injection velocity $v_0 = 3.65$ m/s and the immersion depth $h = 3$ mm) and (b) column-jet regime (The diameter of the needle $d = 0.9$ mm, the inclined angle of needle $\theta = 40^{\circ}$, the injection velocity $v_0 = 3.65$ m/s and the immersion depth $h = 3$ mm). Scale bar for both cases represents 10 mm. }
\label{vortex2}
\end{figure*}

The transition between the curtain-jet regime and the column-jet regime can be understood through the interplay between inertial forces and gravitational forces. This relationship is captured by the Froude number, defined as
\begin{equation}
\label{eqn2}
Fr = \frac{v_0}{\sqrt{g l_0}},
\end{equation}
where $g$ is the gravitational acceleration and $l_0$ is a characteristic length. The characteristic length $l_0$ can be approximated by the maximum height a jet following a parabolic trajectory could reach, given by
\begin{equation}
\label{eqn3}
l_0 = \frac{(v_0 \sin\theta)^2}{2g}.
\end{equation}
Substituting this into the Froude number equation, the ratio of flow inertia to gravitational force simplifies to $Fr = \sqrt{2}/\sin\theta$.  

In this context, the formation of the jet structure is heavily influenced by this balance of forces. When the jet forms a curtain-like structure just above the free surface, a high inertial force combined with a relatively low gravitational force causes the curtain to become broader and thinner. However, as the curtain becomes thinner, it grows increasingly unstable due to flow perturbations, eventually breaking apart and transitioning into a column flow structure.

Experimental observations, as shown in Figure~\ref{regime}, indicate that this transition from the curtain-jet regime to the column-jet regime is primarily governed by the ejection angle, independent of the jet’s Reynolds number. The critical angle at which the column jet forms is found to be approximately $\theta \sim 40^{\circ}$. This corresponds to a critical Froude number of approximately 2.2, beyond which the flow transitions from a curtain structure to a column-like structure. Thus, the key factor driving this transition is the ratio of inertial forces to gravitational forces, as captured by the Froude number, with the ejection angle acting as the primary control parameter in determining whether the jet will form a curtain or column regime. 

However, this conclusion appears to be different from our previous work on laminar, viscous jets~\cite{zou2017pof}. Actually, the transition of a liquid jet from a curtain to a column regime, as investigated in these two studies, is governed by fundamentally different primary parameters due to the distinct flow regimes they examine. The work by Zou et al.~\cite{zou2017pof} focuses on low-Reynolds number, viscous jets, where the transition is primarily velocity-driven. An increase in the jet's exit velocity directly leads to a reduction in the contact length between the jet and the bulk liquid. This occurs because higher velocities reduce the thickness of the entrained viscous boundary layer, allowing the jet to penetrate the free surface more cleanly. Consequently, the jet's morphology evolves from a broad, adhering curtain at low speeds to a detached column as inertia overcomes the capillary forces that bind the jet to the surface. In stark contrast, we study turbulent jets and identify the inclination angle ($\theta$) as the sole critical parameter for transition, independent of the Reynolds number. The underlying physical mechanisms responsible for the transition are also distinct, reflecting the different dominant forces in viscous versus turbulent flow. These two studies present complementary perspectives on jet behavior. In viscous jets, the key mechanism is the competition between capillary force and inertia, mediated by the entrainment of ambient liquid in the tank. In the curtain regime, capillary forces dominate, dragging the jet downwards and shaping it into a sheet-like structure. The transition to the column regime is marked by inertial forces becoming dominant, which minimizes entrainment and severs the extensive capillary connection to the bulk liquid, resulting in a compact column. For the turbulent jets, the transition is a result of the balance between inertial and gravitational forces. The inclination angle directly controls the component of gravity acting normal to the jet's path. At low angles, this balance fosters a broad, fan-shaped curtain that interacts strongly with the pool's surface. Beyond the critical angle, gravity facilitates the detachment of the jet's lower edge, and inertia projects it forward as a discrete column, with surface tension playing a less decisive role.

\subsection{Dynamics of the liquid jet}

To gain deeper insight into the evolution of the jet, red ink was introduced into the flow, and detailed observation experiments were conducted using a high-resolution color camera, as shown in Figure~\ref{vortex2}. Unlike the behavior observed in viscous jets, the high Reynolds number jet demonstrated pronounced turbulence characteristics. In experiments involving high-viscosity jets, the liquid ejected at high velocity from the needle formed well-defined layers, showing clear stratification between the injected liquid and the liquid from the reservoir \cite{zou2017pof}. In particular, in the curtain-jet configuration, the ink-laden liquid primarily concentrated at the uppermost portion of the curtain, creating a distinct flow pattern.

Our investigation revealed that the water jet not only exhibited Kelvin-Helmholtz (K-H) instabilities\cite{Helmholtz1858instability}—caused by high-velocity shear within the submerged part of the jet—but also experienced significant mixing with the reservoir liquid above the water surface. This mixing process resulted in the red ink spreading visibly throughout the entire jet structure, coloring the entire flow. Interestingly, some sections of the jet did not follow the expected outward trajectory (see the snapshot of 64 ms). Instead, these sections diffused and gradually recirculated back into the reservoir, particularly at the point where the curtain jet met the water surface. This occurred as the jet’s velocity diminished, allowing for this recirculatory motion.

\begin{figure}[!t]
\centering
\includegraphics[width = 0.45\textwidth]{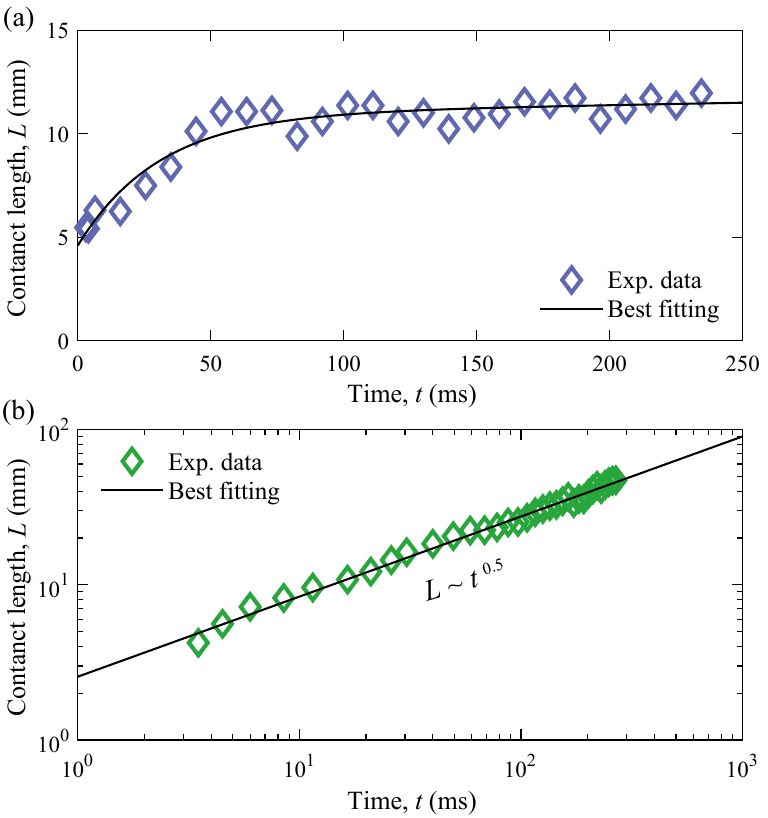}
\caption{Temporal evolution of the contact length $L$ for (a) column-jet regime and (b) curtain-jet regime.}
\label{figtime}
\end{figure}

\begin{figure}[!t]
\centering
\includegraphics[width = 0.45\textwidth]{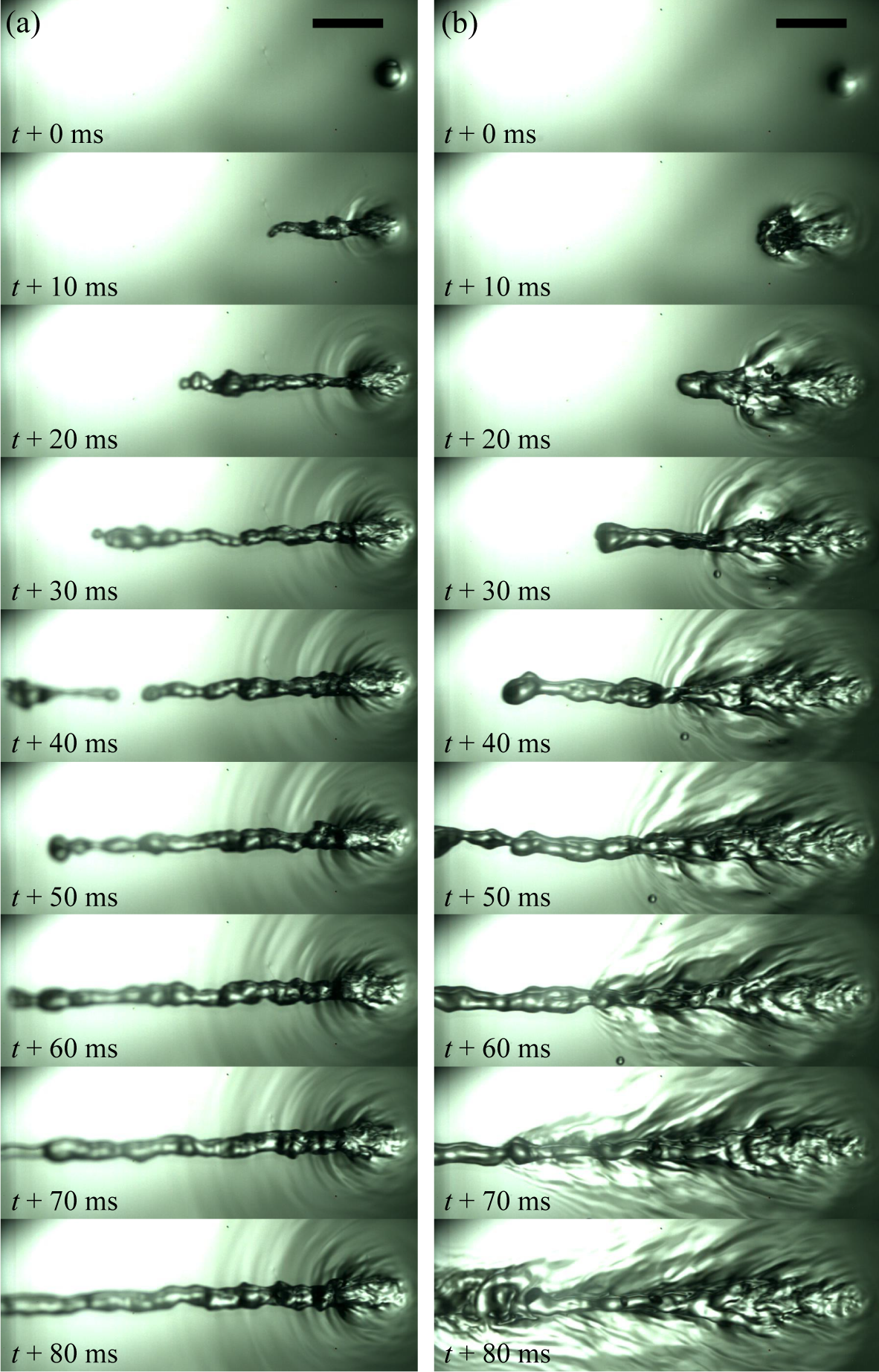}
\caption{Temporal evolution of the leaping jet for (a) column-jet regime (The diameter of the needle $d = 0.9$ mm, the inclined angle of needle $\theta = 40^{\circ}$, the injection velocity $v_0 = 3.65$ m/s and the immersion depth $h = 3$ mm) and (b) curtain-jet regime (The diameter of the needle $d = 0.9$ mm, the inclined angle of needle $\theta = 30^{\circ}$, the injection velocity $v_0 = 3.65$ m/s and the immersion depth $h = 3$ mm). Scale bar for both cases represents 10 mm. }
\label{vortex3}
\end{figure}

In contrast, the column-jet configuration did not display this direct backflow of liquid from the jet into the reservoir (see Figure~\ref{vortex2}(b)). However, due to the intense turbulence present on the surface of the column jet, noticeable undulations and fluctuations were observed along its length. These surface irregularities were a result of the strong shear forces acting between the jet and the surrounding fluid, creating dynamic and unstable flow features throughout the jet's trajectory.

Figure~\ref{figtime} illustrates the temporal evolution of the contact length (between the jet and the water pool) for two distinct regimes: the column jet and the curtain jet. For the column jet, the contact length exhibits a rapid initial increase until it reaches a plateau, after which it stabilizes at a constant value. This behavior suggests that the column jet quickly establishes a consistent interaction with the surrounding fluid, allowing for efficient liquid ejection and entrainment. In contrast, the curtain jet demonstrates a different temporal evolution, following a power law relationship expressed as $L \sim t^{0.5}$. This indicates that the contact length increases with the square root of time, reflecting a more gradual and ongoing process. Unlike the column jet, the contact length for the curtain jet does not quickly reach a constant value, signifying a sustained dynamic interaction with the liquid pool. This prolonged increase in contact length also highlights the continuous variation in the amount of liquid that the curtain jet entrains from the pool. As the curtain jet travels through the submerged layer, it interacts with the surrounding fluid in a way that allows it to draw in additional liquid over time. The power law behavior suggests that the entrainment process is influenced by the jet's geometry and the surrounding hydrodynamics, resulting in a more complex and evolving flow structure compared to the more stable behavior of the column jet. Note that after a long enough time, the curtain jet will also tend to stabilize, i.e., the contact length tends to be constant. 

From a top view, the development of the jet reveals distinct patterns that further illustrate the differences between the column and curtain jets, as shown in Figure~\ref{vortex3}. Initially, the column jet appears as a narrow, well-defined stream, maintaining a cylindrical shape as it exits the pool (Figure~\ref{vortex3}(a)). The uniformity of its flow contributes to a focused jet trajectory, allowing for efficient ejection of liquid and minimal dispersion. In contrast, the curtain jet displays a broader, more fan-shaped structure, characterized by a more complex interplay of vortices and surface waves (Figure~\ref{vortex3}(b)). As it evolves, the curtain jet fans out, creating a dynamic boundary that interacts extensively with the surrounding fluid. This interaction leads to the formation of eddies and turbulent structures, which enhance the entrainment of air and liquid. The top view also highlights the continuous expansion of the contact area for the curtain jet, indicating an ongoing process of liquid capture from the pool. 

\subsection{Critical condition for bubble entrainment} 

\begin{figure}[t!]
\centering
\includegraphics[width = 0.46\textwidth]{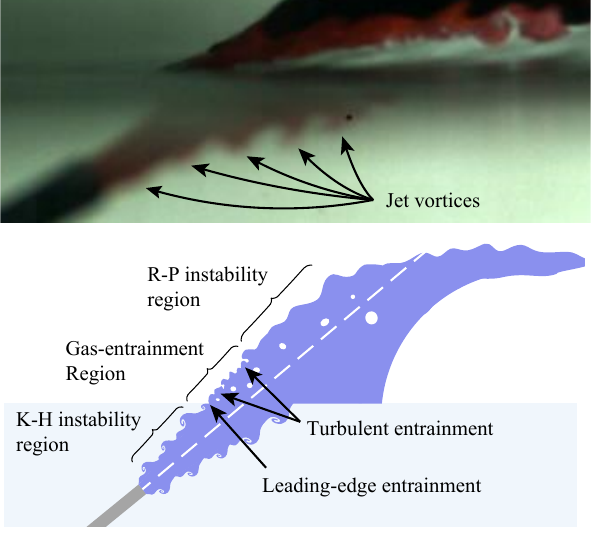}
\caption{Experimental observations and schematics of a side view of the underwater jet shooting out of the water. The water ejected from the needle was marked with red dye. }
\label{vortex666}
\end{figure}

When the water jet exits with sufficient velocity, turbulent disturbances near the point of exit give rise to localized low-pressure zones, which are capable of trapping air bubbles within the jet, as depicted in Figure~\ref{vortex666}. The whole jet structure can be divided into three regions: K-H instability region, gas-entrainment region and R-P instability region. This phenomenon is more pronounced at high Reynolds numbers, where the submerged flow becomes increasingly turbulent, leading to the development of three-dimensional motions that span a wide range of length scales. These motions evolve during the lifetime of the K-H instability, where smaller-scale eddies form and intensify the mixing and stirring within the jet. The presence of these eddies enhances the complexity of the flow and creates pockets of low pressure where air can be entrained. Upon entering an air domain, the stability of a jet is characterized by the Weber number (air phase), representing the ratio of inertial forces to surface tension forces. A Weber number greater than one indicates inertial dominance, where surface tension is insufficient to dampen shear-induced disturbances, resulting in Helmholtz instability. In contrast, a Weber number less than one signifies surface tension dominance, which suppresses disturbances and promotes the onset of Rayleigh instability over Helmholtz instability \cite{eggers2008rpp}. With a Weber number below one in the air, the K-H instability disturbances induced in the water were smoothed out by surface tension. This led to the final breakup of the jet into droplets via R-P instability.

As the jet emerges from the liquid surface, the interaction between the jet and the surrounding air triggers additional entrainment of air, primarily due to the fluctuating pressure fields that develop in the turbulent boundary layer. These pressure fluctuations are a result of the chaotic motion within the jet, where variations in velocity and vorticity at the interface between the water and air create zones of low pressure that can trap and pull air into the flow. Once air bubbles are entrained, they disrupt the previously smooth structure of the jet. The presence of these bubbles introduces additional disturbances, which in turn amplify the existing instabilities within the jet. The bubbles increase the irregularity of the flow, causing the jet to lose coherence more rapidly. This leads to further fragmentation of the jet and enhances the development of secondary instabilities, reinforcing the turbulent nature of the jet and accelerating its breakup into smaller droplets. 

Complementary experiments were conducted to explore how various physical parameters influence bubble entrainment in water jets. Initially, we kept the nozzle diameter and immersion distance constant (with values of $d = 1.45$ mm and $l = 6$ mm, respectively) while increasing both the injection velocity $v_0$ and the inclination angle of the needle $\theta$. The experimental results revealed a strong correlation between bubble entrainment and the Reynolds number, which directly depends on the injection velocity $v_0$. The critical injection velocity for bubble entrainment, denoted as $v_0^* = 3.6$ m/s, remained the same regardless of whether the flow was in the curtain-jet or column-jet regime. This indicates that changes in the overall flow structure have minimal influence on the entrainment of bubbles.

Furthermore, we observed that the immersion depth $h$ increases as the inclination angle increases, while maintaining the immersion distance $l$ constant. The immersion depth can be expressed as $h = l\sin\theta - 1/2 d\cos\theta$. During the process of the underwater jet traversing through the submerged liquid pool layer, it exerts shear forces on the surrounding fluid. These shear forces accelerate the neighboring liquid, causing it to move with the jet and eventually cross the free surface. Consequently, the total flow rate of liquid ejected from the free surface becomes greater than the flow rate at the jet outlet. The amount of liquid entrained by the jet as it crosses the free surface is largely dependent on the immersion distance $l$. A larger immersion distance results in a greater area of shear interaction, leading to higher entrainment of liquid and subsequent ejection. Additionally, due to the conservation of momentum, increased liquid entrainment leads to a reduction in the jet’s initial velocity as it exits the liquid surface.

\begin{figure}[!t]
\centering
\includegraphics[width = 0.48\textwidth]{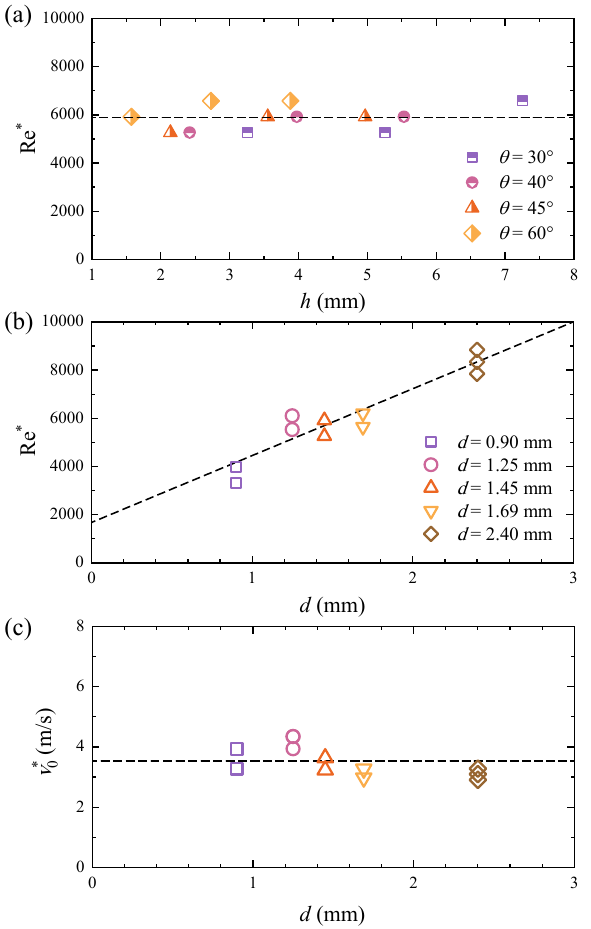}
\caption{(a) The critical Reynolds number $Re^*$ for jets with various immersion distance $h$. (b) The critical Reynolds number $Re^*$ or jets with various nozzle diameter $d$ and a set of immersion depth $h =$ 1, 2 and 3 mm. (c) The dependence between the critical jet velocity $v_0^*$ and the nozzle diameter $d$.}
\label{regime2}
\end{figure}

To further investigate, we performed a series of tests using the same needle (with $d = 1.45$ mm) and varied the immersion distance $l$ by adjusting the inclination angle ($\theta$ = 30$^\circ$, 40$^\circ$, 45$^\circ$ and 60$^\circ$) in combination with different immersion depths $h =$ 1, 2 and 3 mm. The results showed that the immersion parameters, including the distance $l$ and depth $h$, have a relatively minor effect on bubble entrainment (see Figure~\ref{regime2}(a)). Consistent with earlier findings in Figure~\ref{regime} from varying $\theta$, the critical injection velocity for bubble entrainment remained $v_0^* \sim 3.6$ m/s, regardless of changes in immersion depth or inclination angle. While variations in inclination angle and immersion depth can alter the trajectory and shape of the jet (whether the jet forms a curtain or column), they do not significantly affect the key mechanisms responsible for bubble entrainment, such as shear forces, pressure gradients, and turbulence. 

These observations were based on experiments with a constant needle diameter. To determine whether bubble entrainment is governed by a constant Reynolds number or a constant injection velocity, we conducted further tests with needles of different diameters and varying immersion depths $h =$ 1, 2 and 3 mm. The criterion for determining bubble entrainment was primarily based on direct high-speed visualization. The jet velocity was systematically increased from 2.0 m/s in increments of 0.2 m/s to identify the critical velocity at which air entrainment occurred. For each velocity, three independent experiments were performed. To ensure a conservative assessment, a condition was classified as “with bubble entrainment” only when entrainment was consistently observed across all three replicates. The results demonstrated that the critical Reynolds number $Re^*=\rho v_0^* d/\mu$ increased linearly with the inner diameter of the needle, as shown in Figure~\ref{regime2}(b). This finding supports the conclusion that bubble entrainment in water jets is primarily dictated by a constant injection velocity of $\sim$3.6 m/s (see Figure~\ref{regime2}(c)).

The experimental observations thus demonstrate a distinct, velocity-controlled onset of bubble entrainment: once the jet exit velocity exceeds approximately $3.6\ \mathrm{m/s}$, air bubbles begin to be entrained as the jet pierces the free surface. This is why critical entrainment criteria are frequently reported as critical entrainment velocities or onset velocity of air bubble entrainment~\cite{bertola2018physical}, emphasizing the significance of injection velocity in the bubble entrainment process. This finding essentially can be attributed to the interaction between K–H type shear-layer instability and surface-tension–driven interfacial dynamics. As the submerged turbulent jet accelerates upward and approaches the free surface, strong velocity shear between the jet core and the surrounding liquid triggers K–H instabilities. Linear K–H theory (including gravity and surface tension) gives the classical criterion in terms of wave number $k$ as
\begin{equation}
\label{eqn33}
\frac{\rho_1\rho_2}{\rho_1 + \rho_2}(v_1 - v_2)^2 > \frac{\rho_1 g}{k} + \frac{\sigma k}{\rho_1 + \rho_2},
\end{equation}
where the subscripts 1 and 2 represent the two fluids in contact, respectively. Here for a liquid–gas interface ($\rho_1 \gg \rho_2$), the most unstable wave number is $k_m = 2\rho_2 v_2^2/3\gamma$. For small perturbations, surface tension stabilizes short wavelengths, but beyond a threshold velocity, perturbations grow nonlinearly and roll up into vortices. These perturbations amplify rapidly, rolling up into coherent vortical structures whose cores exhibit reduced pressure according to Bernoulli’s principle (i.e. local low pressures of order
the dynamic pressure $P_{\rm low}\sim -{1}/{2}\rho v_{\rm eff}^2$, $v_{eff}$ is the effective velocity or more generally scales like ${1}/{2}\rho v_0^2$). As these vortices migrate toward the interface, they create localized pressure depressions immediately beneath the free surface, deforming it downward. When the deformation becomes sufficiently pronounced, the surface curvature inverts and closes around entrained air, giving rise to bubble formation.

The onset of bubble entrainment can be modeled as a balance between the jet’s dynamic pressure and the free surface’s restoring forces. The jet exerts a pressure ${1}/{2}\rho U^2$ on the interface, while the resistance arises mainly from surface tension $\gamma$ acting through local curvature $\gamma/R$, and to a lesser extent, from the hydrostatic term $\rho g h$ ($h$ is the local cavity depth) associated with local surface depression. The normal stress balance at the deformed interface can be written as
\begin{equation}
\label{eqn333}
-(p-\rho g h) + \frac{1}{2}\rho v^2 = \gamma(\nabla\cdot n),
\end{equation}
where $\gamma (\nabla \cdot n)$ represents the capillary pressure due to surface curvature. Neglecting gravity (since Bond number $Bo=\rho g L^2/\gamma \ll 1$ for millimeter-scale ($L$) cavities), the onset condition for interfacial deformation is given by 
\begin{equation}
\label{eqn336}
\frac{1}{2}\rho v^2 \approx \frac{\gamma}{R}.
\end{equation}
This leads to a characteristic threshold velocity $v_c\sim\sqrt{2\gamma/(\rho R)}$, where $R$ here represents the radius of curvature of the most unstable interfacial perturbation, typically comparable to the shear-layer thickness or the K–H vortex scale. Substituting the physical parameters for water ($\gamma\approx0.072\ \mathrm{N/m}$, $\rho\approx10^3\ \mathrm{kg/m^3}$) and a sub-millimeter–scale $R$ yields $v_c \approx 2 \sim 4\ \mathrm{m/s}$, in agreement with the experimentally observed entrainment threshold. Thus, bubble entrainment occurs when the jet’s dynamic pressure overcomes surface-tension-induced curvature pressure, producing a local cavity that captures air.

Besides, the Weber number provides an auxiliary, nondimensional framework to describe this competition between inertial and capillary effects. Defined as ${We}=\rho v^2 D/\gamma$ ($D$ is the jet diameter), it expresses the ratio of dynamic to surface-tension forces. For jet diameters in the range $D =1 \sim 2$ mm and a velocity $U\approx3.6\ \mathrm{m/s}$, ${We}$ is on the order of $200 \sim 300$, indicating that inertia overwhelmingly dominates surface tension. Empirically, interfacial rupture and air entrainment are known to occur for ${We}\gtrsim10^2$, consistent with this magnitude~\cite{eggers2008physics}. However, while ${We}$ changes with nozzle diameter, the entrainment threshold expressed in terms of velocity does not. This is because the governing condition involves the local pressure balance, ${1}/{2}\rho v^2\approx\gamma/R$, which depends only on $v$ and intrinsic fluid properties, not on macroscopic geometric parameters. The Weber number merely reflects that the flow is in a regime where the interface is deformable and susceptible to rupture, but the threshold itself remains velocity-determined. 

The constancy of the critical velocity $v_0^*\approx3.6\ \mathrm{m/s}$ across different geometries and flow regimes thus arises from the universality of this local dynamic–capillary balance. Altering the inclination angle, immersion depth, or nozzle diameter modifies the jet trajectory, shape, and Reynolds number $Re$, yet these changes do not significantly affect the shear-layer intensity or the local pressure field at the jet–surface intersection (a condition that depends solely on $\rho$, $\gamma$, and $v$). The K–H instability always produces vortical structures of similar characteristic scale near the exit, and the same velocity is required for their dynamic pressure to overcome surface tension. Consequently, the onset of bubble entrainment is controlled by an invariant physical mechanism, the inertial stress needed to breach the capillary barrier, and remains insensitive to external geometric variations. This explains why the entrainment threshold of approximately $3.6\ \mathrm{m/s}$ is globally constant, reflecting a fundamental interplay between shear-induced vortical dynamics and surface-tension-driven interface stability.

\section{Conclusions}

This study provides a comprehensive analysis of water jet behavior, focusing on the transition between flow regimes and the mechanisms driving bubble entrainment. The findings indicate that the transition between the curtain-jet and column-jet regimes is primarily governed by the balance between inertial and gravitational forces. This balance is captured by the Froude number, with a critical injection angle of approximately $\theta = 40^{\circ}$, corresponding to a critical Froude number of around 2.2, marking the shift from curtain flow to column flow. Despite differences in flow structure between the two regimes, the results demonstrated that bubble entrainment is not significantly affected by the transition between these regimes.

The entrainment of bubbles is due to the excitation of underwater jets during their traversal of a thin water layer, caused by K-H instability, which results in the amplification of shear perturbations between the adjacent fluids within the aquatic medium. Subsequently, when the jet transitions into the air domain, these perturbations evolve into the entrainment and formation of bubbles. Therefore, the most crucial factor influencing bubble entrainment is the injection velocity, as shown by a constant critical velocity $\sim3.6$ m/s. This critical velocity is necessary for bubble entrainment, regardless of the flow regime or variations in nozzle inclination and immersion depth. 

This study highlights the critical role of injection velocity in determining bubble entrainment in water jets, irrespective of the flow regime or other physical parameters. The insights gained from this work offer a clearer understanding of the underlying dynamics of water jets and can serve as a foundation for future studies and applications involving bubble entrainment in fluid systems.

\section*{Acknowledgments}
This work was supported by National Key Research and Development Program of China (Grant No. 2022YFB4602502), the Zhejiang Provincial Natural Science Foundation of China (LD22E050003), and National Natural Science Foundation of China under Grants Nos. 12572290 and 12202244.

\section*{Compliance with ethical standards Conflict of interest:}
\textbf{Conflict of interest:} The authors declare that they have no conflict of interest. All authors declare that there are no other competing interests.

\textbf{Ethical approval:} This article does not contain any studies with human participants or animals performed by any of the authors.

\textbf{Informed consent:} Not applicable.

\section*{Data Availability}

The data that supports the findings of this study are available within the article and its supplementary material.

\section*{Author contributions}
\textbf{Fangye Lin:} Conceptualization, Methodology, Writing-original draft, Software, Formal analysis, Funding acquisition. 

\textbf{Mingbo Li:} Visualization, Validation, Writing-review \& editing, Methodology, Supervision

\appendix
\section*{Appendix A. Supplementary Material}

Supplementary data associated with this article can be found in four videos. 

\bibliographystyle{elsarticle-num}



\end{sloppypar}

\end{document}